\documentclass[12pt]{article}
\usepackage{times,graphicx}
\usepackage[super]{natbib} % bibliography
\usepackage{geometry}
\usepackage[utf8]{inputenc}
\usepackage{enumitem,amssymb}
\usepackage{ragged2e}
\newlist{thematic}{itemize}{8}
\setlist[thematic]{label=$\square$}
\usepackage{pifont}
\newcommand{\cmark}{\ding{51}}%
\newcommand{\done}{\rlap{$\square$}{\raisebox{2pt}{\large\hspace{1pt}\cmark}}%
\hspace{-2.5pt}}

\begin{document}
\raggedright
\huge
Astro2020 Science White Paper \linebreak

Compact Stellar Jets \linebreak
\normalsize

\noindent \textbf{Thematic Areas:} \hspace*{60pt} $\square$ Planetary Systems \hspace*{10pt} $\square$ Star and Planet Formation \hspace*{20pt}\linebreak
$\done$ Formation and Evolution of Compact Objects \hspace*{31pt} $\square$ Cosmology and Fundamental Physics \linebreak
  $\square$  Stars and Stellar Evolution \hspace*{1pt} $\square$ Resolved Stellar Populations and their Environments \hspace*{40pt} \linebreak
  $\square$    Galaxy Evolution   \hspace*{45pt} $\square$             Multi-Messenger Astronomy and Astrophysics \hspace*{65pt} \linebreak
  
\textbf{Principal Author:}

Name:	
Thomas J. Maccarone
 \linebreak						
Institution:  
Texas Tech University
 \linebreak
Email: 
thomas.maccarone@ttu.edu
 \linebreak
Phone:  
+1 806 834 3760
 \linebreak
 
 %
 % SH: Would suggest alphabetic order of the co-author list, simply because anything else will deviate from other white papers
 %
 
\justify 
 
\textbf{Co-authors:} %(names and institutions)
Elena Gallo (University of Michigan), Sebastian Heinz (University of Wisconsin-Madison), James C. A. Miller-Jones (Curtin University), Piergiorgio Casella (INAF-OAR), Stephen Eikenberry (University of Florida), Poshak Gandhi (University of Southampton),  Richard M. Plotkin (University of Nevada), Gregory R. Sivakoff (University of Alberta), James F. Steiner (MIT), Alexandra J. Tetarenko (East Asian Observatory), John A. Tomsick (UC Berkeley/SSL)
  \linebreak

\textbf{Abstract:}\\
This paper outlines the importance of understanding jets from compact binaries for the problem of understanding the broader phenomenology of jet production.  Because X-ray binaries are nearby and bright, have well-measured system parameters, and vary by factors of $\sim 10^6$ on $\sim$ year timescales, they provide a unique opportunity to understand how various aspects of the jet physics change in response to changes in the accretion flow, giving the possibility of looking for trends within individual systems and testing their universality with other systems, rather than trying to interpret large samples of objects on a statistical basis.

\pagebreak
%Insert your white paper text here (max of five pages including figures).
\begin{center}{\bf\Large Jets from Compact Stellar Accretors}
\end{center}
\vspace{1pt}

{\noindent Jets are of great importance across a broad range of topics in astrophysics.  The jets from active galactic nuclei are one of the primary sources of kinetic energy feedback, which strongly affects cosmic structure formation.  The motivation to understand jets has gained new urgency in the age of multi-messenger astronomy, most importantly through the realization that relativistic jets from blazars produce at least some of the most energetic cosmic neutrinos \citep{icecube:18} and are thus a promising source of ultra-high energy cosmic rays.  While some work can be done to understand the process of jet production in active galactic nuclei, the use of X-ray binaries, with their more rapid variability, has provided the strongest insights into the coupling of accretion and jet production to date.  By watching the variability of a single X-ray binary, one controls for the mass and the angular momentum of the accretor, and can observe how the jet properties change as a function of the accretion rate and the accretion rate history.}

Furthermore, there are other types of physics that can be extracted more readily from X-ray binary jets than from active galactic nucleus jets.  One expects to see correlations of the jet's intrinsic power on properties like the accretion rate, mass, and spin of the accretor, and also to see the effects of parameters like inclination angle.  Masses are usually more reliably measured for X-ray binary accretors than for supermassive black holes.  For spins, the question remains murkier -- in principle, spectral methods based on the blurring of fluorescence features from the accretion disks should be equally effective for estimating black hole spins for stellar mass and supermassive black hole accretors. An advantage of the stellar mass black holes is that there are two different methods--fluorescence spectroscopy and continuum fitting of the accretion disks' thermal spectra--that are in widespread usage, and a third method, the use of high frequency quasi-periodic oscillations that is subject to clear modelling degeneracies, but that gives precise measurements subject to the model interpretation.  At the present time, the different methods do not agree with one another in all cases, so more work is needed to disentangle the physics, but because of the ability to cross-check answers, there is greater hope for the stellar black holes to give reliable spin measurements in the near future than for the supermassive black holes to do so.  Importantly, the canonical method of spin measurements in AGN, relativistic iron line fitting, is not accessible for typical radio loud AGN, which are in a different accretion state. X-ray binary jets thus offer the only reliable way to test the dependence of jet production on spin.

The central hypothesis of jet production is that, for black holes, the processes responsible for launching jets are scale invariant, so that jets from black holes of different masses and horizon sizes share a common structure and dynamical properties, all other non-dimensional parameters (accretion rate, spin, magnetic flux) being equal \citep[]{heinz:03,merloni:03,falcke:04}. This suggests that we can use the more easily observable properties of jets from stellar mass black holes to inform our understanding of jets even in the most massive AGN believed to be responsible for cosmic feedback. Furthermore, with accreting compact stars, the properties of accreting black holes can be compared with those of accreting neutron stars to determine which effects are generic to accretion, which require relativistic gravitational potentials, and which are the results of the presence of a surface and perhaps a magnetic field.  Here, the neutron star spins are often directly measurable through other means.

{\noindent\bf Phenomenology of jet power production}

Over the late 1990s and early 2000s a general picture developed for the production of jets in accreting black holes and neutron stars.  A few general themes emerged:\citep[]{fenderbook} 
\begin{itemize}[noitemsep]
\item Jet radio emission is stronger for higher hard X-ray power.
\item Jet radio emission becomes undetectable (likely turns off) for black holes in ``high/soft'' states, where the accretion disk is geometrically thin and optically thick, rather than in ``hard states'' where there is some sort of geometrically extended, optically thin or translucent Comptonizing region.
\item Jets from black hole disks are more powerful than those from neutron star disks in the same spectral states with the same X-ray luminosity.
\item High magnetic field neutron stars, where the accretion disk is disrupted by the magnetosphere of the neutron star, show significantly fainter radio emission.
\item ``Ballistic'' jets appears when sources make strong state transitions (typically at $\approx$10\% or more of the Eddington luminosity), while ``steady'' jets appear when sources remain steadily in low/hard states.
\end{itemize}

\begin{figure}[t]
    \centering
    \includegraphics[height=2.6in]{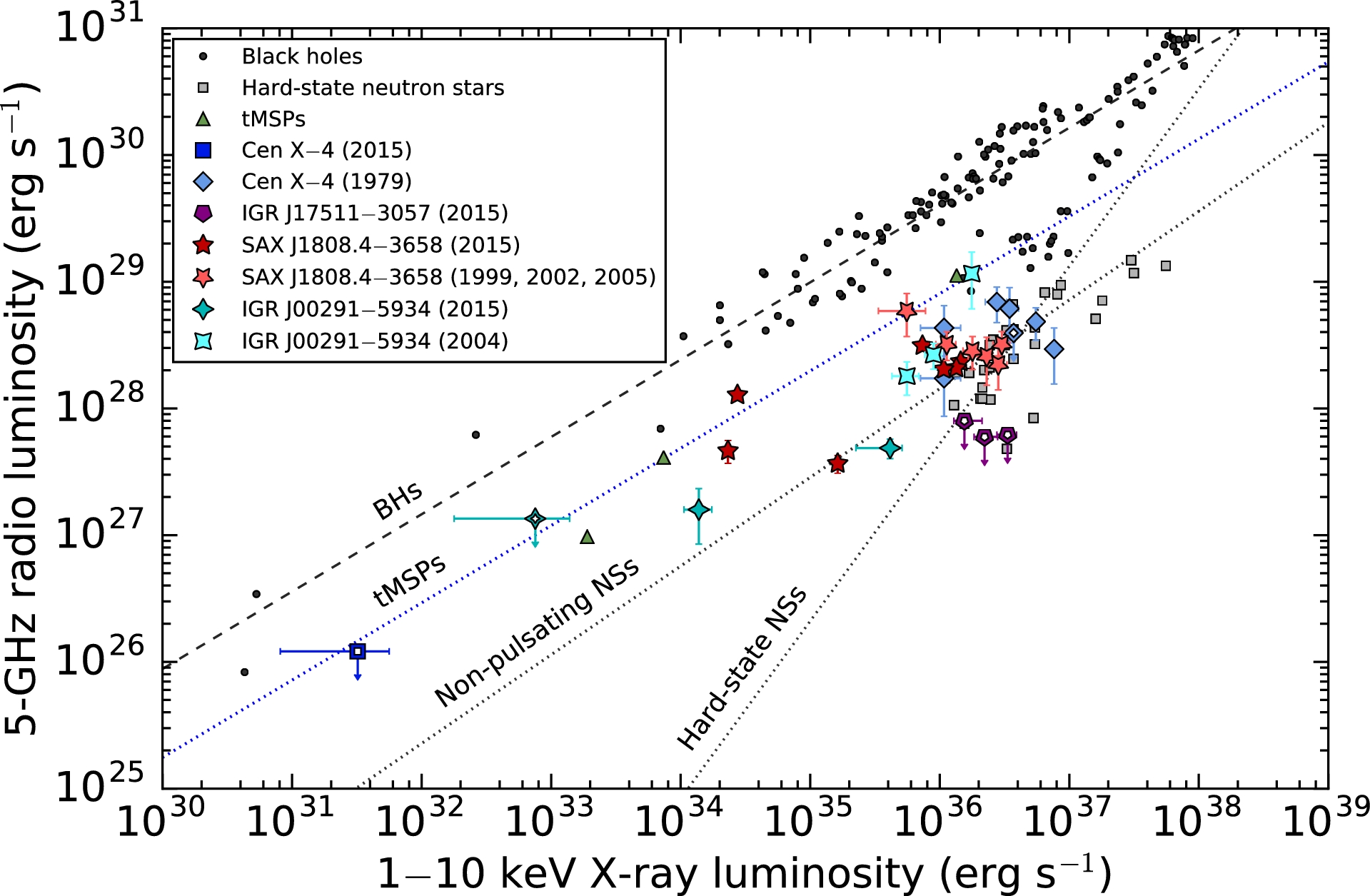} \includegraphics[height=2.6in]{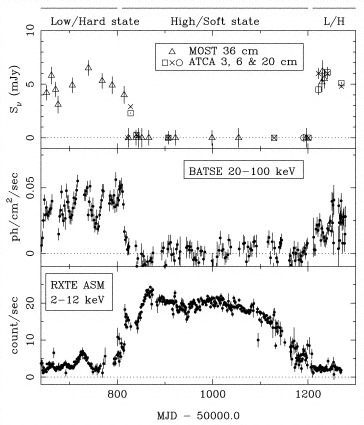}
    \caption{Left: Radio luminosity plotted versus X-ray luminosity for a variety of systems in geometrically thick, optically thin accretion states\cite{tudor17}.  The black dots show the black holes, while the colored symbols are neutron stars. Right: An illustration of the ``quenching'' of jets of X-ray binaries as the X-ray spectra become soft X-ray dominated, and a highlight of the characteristic timescales of X-ray binary outbursts\cite{fender99}.}
    \label{fig:tudor}
\end{figure}

These themes have largely remained valid, although they have been complicated by some recent observational results.  While the early data sets indicated that, for black holes, $L_R \propto L_X^{0.7}$\cite{corbel03,gfp03}, more recent work shows that some fraction of sources show a steeper relation when the sources are bright, perhaps because of geometric beaming effects\cite{coriat11,gallo14,motta18}.  For neutron stars, data quality remains a problem.  It has been well established that the neutron stars are fainter than black holes at similar X-ray luminosities when in the hard states\cite{gallo18} (see Fig.~\ref{fig:tudor}).

It remains uncertain whether the jet powers correlate strongly with the inferred black hole spins.  This problem stems, in part from the uncertainties in the spins themselves, but it also is affected by the diagnostics used.  When steady jets from low power accretors are used, and the diagnostic of the effect of spin is the normalization of the relation between $L_R$ and $L_X$, no correlation is seen\cite{fender10}. When peak flare luminosity is used, a correlation seems to appear\cite{nmc12,russellspin}  The two approaches need not yield the same result, if the coupling of the accretion flow to the black hole's spin behaves differently for the steady and ballistic jets.  Further, potential biases exist for both approaches -- both are based on relatively small samples of sources, with the jet inclination angles not all well known (but with beaming effects potentially important) and the radio luminosity used as a proxy for jet power does not have a 1-to-1 mapping with the kinetic power input into the jet in either ballistic or compact jets.

To make further progress with the ballistic jet ejections requires higher cadence monitoring of these sources when they are bright, something which requires higher quality all-sky monitors than presently exist.  Making further progress on the steady jet sources can be done largely by obtaining better system parameters -- distances, masses, and inclination angles, as well as better diagnostics for jet power. The latter can be obtained either from calorimetric measures (such as the interaction of ballistic jets with the interstellar environment\citep{tetarenkoism}) or instantaneous measures, such as observations of the interaction of a jet with the wind of the companion star, as in Cygnus X-3\citep{corbel12,koljonen18}.

There is reason for significant optimism for major progress on the theoretical front: For the first time since jets were discovered in 1918, we are at the threshold of being able to model their production ab-initio, using GRMHD simulations of black hole accretion flows \citep[]{mckinney:06,liska:18}. This ability should motivate commensurate efforts in milli-arcsecond radio and sub-mm spectro-polarimetry. While we have no hope of resolving the horizon scale for the case of compact stellar remnants, both because of optical depth and lack of resolution, the ability to correlate jet properties on AU scales in real time with accretion rate changes and binary orbit have already produced key insight into jet physics and will continue to be the key enabling technology in this field.

{\noindent\bf Understanding jet acceleration}

In principle, the best way to measure jet acceleration is through direct imaging.  For a small fraction of AGN jets, direct measurements can be made through radio proper motion studies \citep{mojave16}.  For X-ray binary jets, however, the best means available is through timing measurements.
%
% SH: These really probe different scales, so mentioning both is important. While timing probes the right region for acceleration, it is much more ambiguous to interpret. Compared to AGN jets, compact jets from binaries suffer from the fact that the radio emission is optically thick until quite far out, so perhaps the best we can hope for here is understanding acceleration in AGN jets like M87 and then develop diagnostics at other wavelengths that allow a comparison.
%

This can be done by starting from the point that when jets are emitting steadily, with just stochastic variability, they are summations of emission from many different radii.  At each radius, the jet spectrum will be rather strongly peaked, with the low frequency emission lost to synchrotron self-absorption, and the higher frequency emission falling off due to the $F_\nu \propto \nu^{-0.7}$ spectrum typical of synchrotron emission from particles accelerated via the Fermi mechanism.  For jets viewed edge-on with the same total power at each radius, this produced a flat spectrum in $F_\nu$, as typically observed\cite{bk79}, with a break then associated with the synchrotron self-absorption frequency at the bottom of the jet\cite{gandhi11,russell13_maxi1836}.  More realistic models of the jet can have somewhat different energy spectra, but the basic picture is likely to hold in all systems where a broad range of radii contribute to the emission from the jet.

\begin{figure}[t]
    \centering
    \includegraphics[width=2.7in]{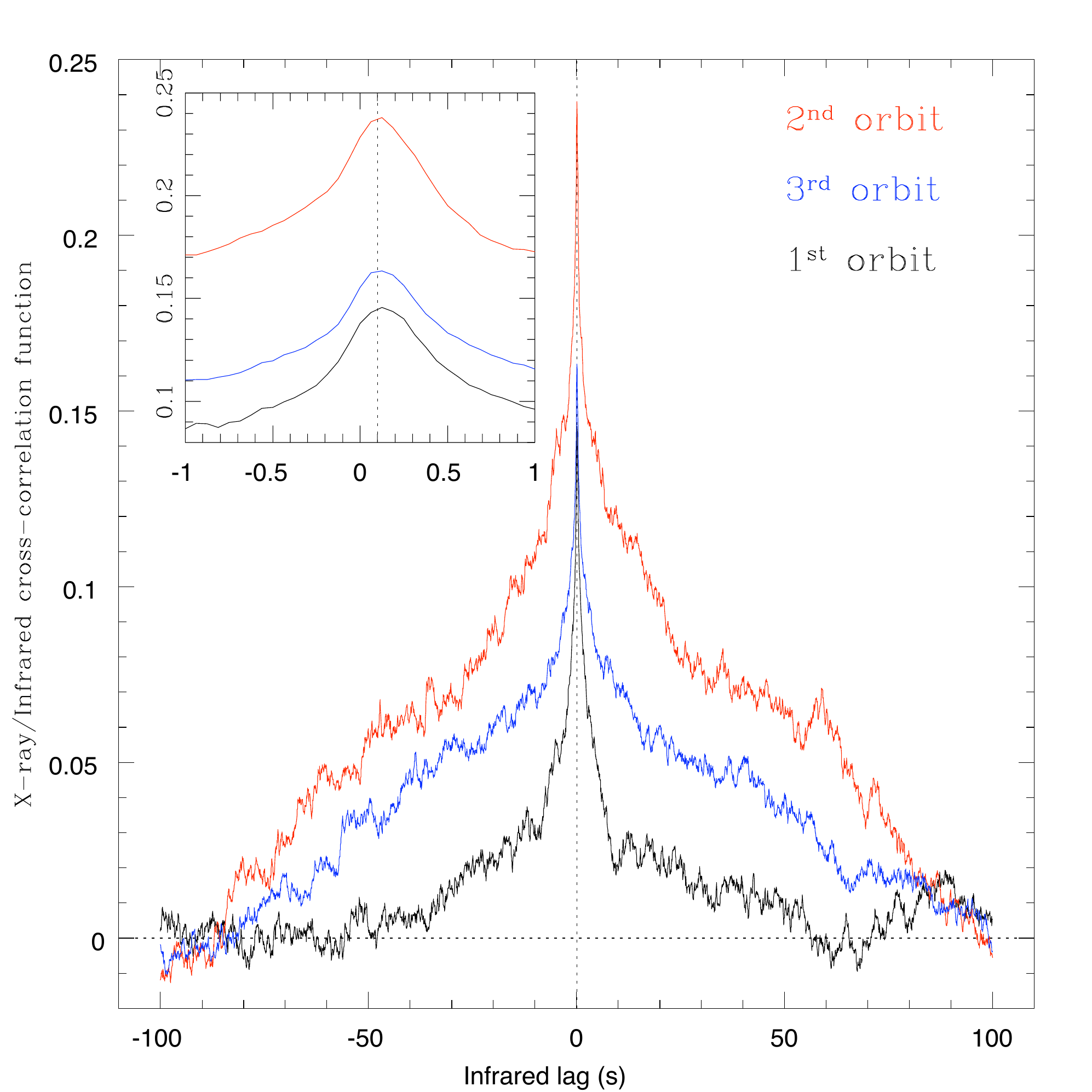}\includegraphics[width=3.7in]{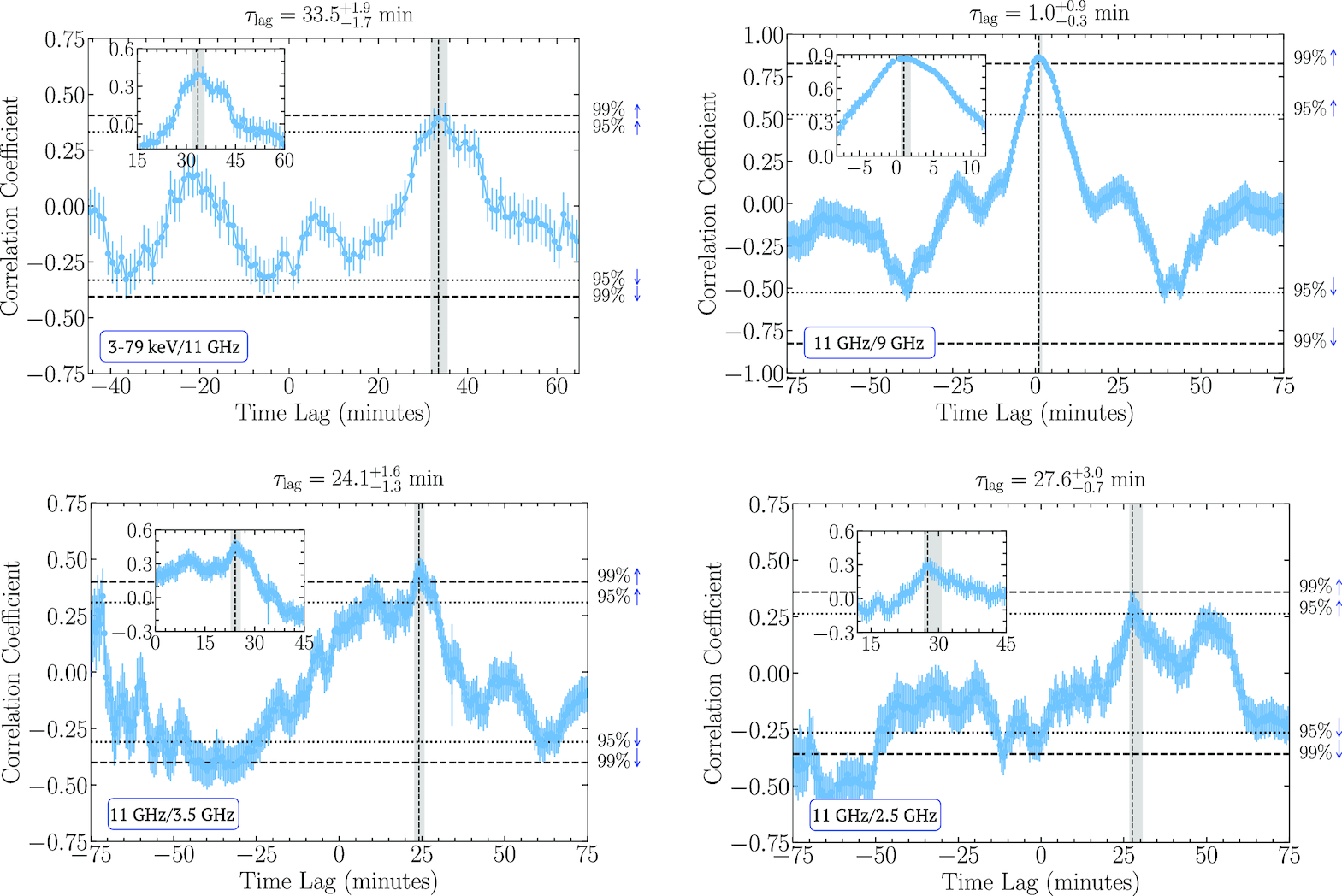}
    \caption{Two different observations of cross-correlation functions of time series from X-ray binaries at different wavebands.  Left: An infrared-versus X-ray cross-correlation function for GX~339-4, showing that the infrared emission, which comes from the jet, lags the X-ray emission, which comes from the accretion disk, by about 0.1 seconds but the two bands are well-correlated.  Right: Cross-correlations among many different sets of bands for Cygnus X-1, indicating significantly longer lags for the radio emission behind the X-ray emission, and furthermore, even longer lags, and more smeared variability at 2 GHz than at 10 GHz for the radio emission.  At the present time, we have obtained good enough data to highlight the efficacy of the technique, but it is necessary to have many more observations of this sort to make actual maps of delay time versus wavelength as a function of the different parameters likely to affect jet acceleration.}
    \label{fig:Acceleration}
\end{figure}

The shortest wavelength bands thus come from the inner parts of the jet, while progressively longer wavelengths come from further from the central engine.  By obtaining time lags between the X-rays and the emission at many other wavelengths, and then having direct measurements of the jet's length in the radio, one can map out how quickly matter moves through the jet and obtain an understanding of its acceleration on scales from the base at $\sim10^8$ cm to the radio emitting region at $\sim10^{14}$ cm (see Fig.~\ref{fig:Acceleration}).  At the present time, this work has been established to be of value with measurements of a few systems at a few wavelengths\cite{casella10,tetarenko19}.  What is needed to move forward is to have a broad set of wavelengths obtained simultaneously on many occasions, spanning a range of sources and a range of luminosities. Coordinated rapid multiwavelength observations straddling state transitions can be  particularly valuable for tracing changes in the inner jet\cite{gandhi17}.

%{\bf The ULX-microquasar connection}

{\noindent\bf Understanding neutron star and white dwarf jets}

Regardless of whether the black hole spin is {\it important} for accelerating highly relativistic and/or powerful jets from accreting black holes, an ergosphere cannot be a requirement for producing jets, demonstrated by the presence of jets coming from accretion disks around other classes of objects, like neutron stars, white dwarfs, and young stellar objects.  The YSO jets are usually thermal in nature and may represent a different phenomenon from those seen around compact objects, however, jets in particular from neutron stars appear similar in most of their key properties to those accelerated by black holes.  The study of white dwarf jets also holds great promise, since in these systems relativistic effects are negligible, although the jets from these systems are quite faint\citep{coppejans16}, and so new, more sensitive radio facilities are needed to go beyond mere detection into the regime of being able to make correlations between jet power and other parameters.  Having the combination of black holes, neutron stars and white dwarfs with which to work allows the development of comparisons among various factors that may be important such as compact object spin, depth of the gravitational potential well, and magnetic field of the accretor. 

%The inability to directly measure bulk speeds in compact non-thermal jets (with a very small number of notable exceptions) complicates statements of similarity between different source classes. Re-doubled efforts to search for thermal line emission from jets with more sensitive X-ray telescopes provide one way to break this degeneracy, but direct proper motion measurement, and the sorts of timing measurements discussed above can also be applied to the brightest neutron star binaries. 
% Note from Tom: I commented this out because of space reasons, and also because of a concern that whatever jets show emission lines will have likely entrained a lot of material and not be terribly representative

A key advantage of neutron star jets is that their spins can often be directly measured via pulsations, either due to magnetically channeled accretion, or due to asymmetries in the temperatures of emission during thermonuclear bursts.  A high collecting area X-ray detector system would make it likely that even more such pulsations could be measured.  Furthermore, the neutron stars' magnetic fields may be important in ways that are quite complicated, but this, too, can be studied.  It appears that the radio emission from the moderately magnetic accreting millisecond pulsars is significantly stronger than for the un-pulsed accreting neutron stars\citep{deller15}.  On the other hand, the slow accreting pulsars, with much higher magnetic fields that truncate the disks far from the stars' surfaces, have just barely been detected in the radio for an extremely bright X-ray source\cite{eijnden18}.  As a result, it seems likely that moderate magnetic fields help seed the jet formation process, but only when the accretion flow can still extend deep into the gravitational potential well where it rotates quickly, but clearly much larger samples of objects and numbers of epochs per source are needed to make definitive statements.

%Perhaps the most promising avenue for understanding the ability of different accreting compact objects to produce jets will once again come from GRMHD simulations, which are advancing to the point of putting the ab-initio simulation of the inner accretion flow within reach.

%{\bf Spectral energy distributions of jets}
%EG: we need a paragraph about jet power and spin parameter uncertainty (happy to do that); plus one on jet breaks/SEDs (again happy to contribute -- OK -- I think that we also very much need to emphasize the things that point toward new relevant facilities!!  I agree.  I think we may have room for both, but let's see what happens.  I think getting rid of the ULXs will have to happen, though.).

{\noindent \bf Observational capabilities needed to move forward}

Accreting stellar compact objects are especially observationally demanding sets of transients. They are relatively rare, so each one is important to observe intensively, and they vary substantially on timescales from milliseconds to months.  It is thus important to have flexible scheduling (both from a management point of view and from the point of view of having the ability to observe large fractions of the sky at any given time), and to have the capability of observing very bright sources without problems due to pile-up, other forms of saturation or telemetry limitations.  Sensitive wide field monitors in the X-rays are essential both for knowing when transients have taken place and for having well-sampled light curves for comparison with data at other bands.  For rapid variability studies, it is additionally important to have systems with time resolution of at least 10 milliseconds across radio, IR, optical and X-rays, and to have the ground-based instrumentation with multiple channels obtained at a time through the use either of dichroics and filters, or energy sensitive OIR detectors like microwave kinetic induction devices.   Furthermore, all these facilities need rapid response to transients and scheduling coordination with other facilities.

For the studies of the correlations between X-ray properties and radio properties on day to month timescales, larger samples of objects are needed, but the quality of the individual data sets produced now are generally adequate.  In large part, then progress depends on maintaining current observational capabilities. For the accreting neutron stars and white dwarfs, on the other hand, more sensitive radio facilities are needed.

Additionally, many aspects of understanding X-ray binary jets require better system parameters.  Geometric parallax distances are vital in their own right, and also because they can help break degeneracies in estimates of the black holes' masses and spins\citep{Gou11}.  Optical parallaxes are being measured by the {\it Gaia} mission\cite{gandhi19}, but further observations are needed to improve upon current constraints.  The sample of objects with good distance estimate could be increased by a few via a bandwidth upgrade to the current Very Long Baseline Array, but for routine geometric parallax measurements to be made for the typically $\sim10$ microJansky quiescent sources will require something like the Next Generation Very Large Array.  For improving our understanding of the black hole spins, high collecting area X-ray missions are vital (see e.g. the white paper by Javier Garcia).

\pagebreak
\subsection*{References}
{
\renewcommand{\section}[2]{}
\bibliographystyle{mn2e}
\bibliography{bibxrb}

\begin{thebibliography}{33}
\expandafter\ifx\csname natexlab\endcsname\relax\def\natexlab#1{#1}\fi

\bibitem[{{Blandford} \& {Znajek}(1977)}]{bk79}
{Blandford} R.~D., {Znajek} R.~L., 1977, \mnras, 179, 433

\bibitem[{{Casella} {et~al}\mbox{.}(2010){Casella}, {Maccarone}, {O'Brien},
  {Fender}, {Russell}, {van der Klis}, {Pe'Er}, {Maitra}, {Altamirano},
  {Belloni}, {Kanbach}, {Klein-Wolt}, {Mason}, {Soleri}, {Stefanescu},
  {Wiersema}, \& {Wijnands}}]{casella10}
{Casella} P. {et~al.}, 2010, \mnras, 404, L21

\bibitem[{{Coppejans} {et~al}\mbox{.}(2016){Coppejans}, {K{\"o}rding},
  {Miller-Jones}, {Rupen}, {Sivakoff}, {Knigge}, {Groot}, {Woudt}, {Waagen}, \&
  {Templeton}}]{coppejans16}
{Coppejans} D.~L. {et~al.}, 2016, \mnras, 463, 2229

\bibitem[{{Corbel} {et~al}\mbox{.}(2012){Corbel}, {Dubus}, {Tomsick},
  {Szostek}, {Corbet}, {Miller-Jones}, {Richards}, {Pooley}, {Trushkin},
  {Dubois}, {Hill}, {Kerr}, {Max-Moerbeck}, {Readhead}, {Bodaghee}, {Tudose},
  {Parent}, {Wilms}, \& {Pottschmidt}}]{corbel12}
{Corbel} S. {et~al.}, 2012, \mnras, 421, 2947

\bibitem[{{Corbel} {et~al}\mbox{.}(2003){Corbel}, {Nowak}, {Fender},
  {Tzioumis}, \& {Markoff}}]{corbel03}
{Corbel} S., {Nowak} M.~A., {Fender} R.~P., {Tzioumis} A.~K., {Markoff} S.,
  2003, \aap, 400, 1007

\bibitem[{{Coriat} {et~al}\mbox{.}(2011){Coriat}, {Corbel}, {Prat},
  {Miller-Jones}, {Cseh}, {Tzioumis}, {Brocksopp}, {Rodriguez}, {Fender}, \&
  {Sivakoff}}]{coriat11}
{Coriat} M. {et~al.}, 2011, \mnras, 414, 677

\bibitem[{{Deller} {et~al}\mbox{.}(2015){Deller}, {Moldon}, {Miller-Jones},
  {Patruno}, {Hessels}, {Archibald}, {Paragi}, {Heald}, \&
  {Vilchez}}]{deller15}
{Deller} A.~T. {et~al.}, 2015, \apj, 809, 13

\bibitem[{{Falcke}, {K{\"o}rding} \& {Markoff}(2004){Falcke}, {K{\"o}rding}, \&
  {Markoff}}]{falcke:04}
{Falcke} H., {K{\"o}rding} E., {Markoff} S., 2004, \aap, 414, 895

\bibitem[{{Fender}(2006)}]{fenderbook}
{Fender} R., 2006, {Jets from X-ray binaries}, {Lewin} W.~H.~G., {van der Klis}
  M., eds., pp. 381--419

\bibitem[{{Fender} {et~al}\mbox{.}(1999){Fender}, {Corbel}, {Tzioumis},
  {McIntyre}, {Campbell-Wilson}, {Nowak}, {Sood}, {Hunstead}, {Harmon},
  {Durouchoux}, \& {Heindl}}]{fender99}
{Fender} R. {et~al.}, 1999, \apjl, 519, L165

\bibitem[{{Fender}, {Gallo} \& {Russell}(2010){Fender}, {Gallo}, \&
  {Russell}}]{fender10}
{Fender} R.~P., {Gallo} E., {Russell} D., 2010, \mnras, 406, 1425

\bibitem[{{Gallo}, {Degenaar} \& {van den Eijnden}(2018){Gallo}, {Degenaar}, \&
  {van den Eijnden}}]{gallo18}
{Gallo} E., {Degenaar} N., {van den Eijnden} J., 2018, \mnras, 478, L132

\bibitem[{{Gallo}, {Fender} \& {Pooley}(2003){Gallo}, {Fender}, \&
  {Pooley}}]{gfp03}
{Gallo} E., {Fender} R.~P., {Pooley} G.~G., 2003, \mnras, 344, 60

\bibitem[{{Gallo} {et~al}\mbox{.}(2014){Gallo}, {Miller-Jones}, {Russell},
  {Jonker}, {Homan}, {Plotkin}, {Markoff}, {Miller}, {Corbel}, \&
  {Fender}}]{gallo14}
{Gallo} E. {et~al.}, 2014, \mnras, 445, 290

\bibitem[{{Gandhi} {et~al}\mbox{.}(2017){Gandhi}, {Bachetti}, {Dhillon},
  {Fender}, {Hardy}, {Harrison}, {Littlefair}, {Malzac}, {Markoff}, {Marsh},
  {Mooley}, {Stern}, {Tomsick}, {Walton}, {Casella}, {Vincentelli},
  {Altamirano}, {Casares}, {Ceccobello}, {Charles}, {Ferrigno}, {Hynes},
  {Knigge}, {Kuulkers}, {Pahari}, {Rahoui}, {Russell}, \& {Shaw}}]{gandhi17}
{Gandhi} P. {et~al.}, 2017, Nature Astronomy, 1, 859

\bibitem[{{Gandhi} {et~al}\mbox{.}(2011){Gandhi}, {Blain}, {Russell},
  {Casella}, {Malzac}, {Corbel}, {D'Avanzo}, {Lewis}, {Markoff}, {Cadolle Bel},
  {Goldoni}, {Wachter}, {Khangulyan}, \& {Mainzer}}]{gandhi11}
{Gandhi} P. {et~al.}, 2011, \apjl, 740, L13

\bibitem[{{Gandhi} {et~al}\mbox{.}(2018){Gandhi}, {Rao}, {Johnson}, {Paice}, \&
  {Maccarone}}]{gandhi19}
{Gandhi} P., {Rao} A., {Johnson} M.~A.~C., {Paice} J.~A., {Maccarone} T.~J.,
  2018, MNRAS in press, arXiv:1804.11349

\bibitem[{{Gou} {et~al}\mbox{.}(2011){Gou}, {McClintock}, {Reid}, {Orosz},
  {Steiner}, {Narayan}, {Xiang}, {Remillard}, {Arnaud}, \& {Davis}}]{Gou11}
{Gou} L. {et~al.}, 2011, \apj, 742, 85

\bibitem[{{Heinz} \& {Sunyaev}(2003)}]{heinz:03}
{Heinz} S., {Sunyaev} R.~A., 2003, \mnras, 343, L59

\bibitem[{{IceCube Collaboration} {et~al}\mbox{.}(2018){IceCube Collaboration},
  {Aartsen}, {Ackermann}, {Adams}, {Aguilar}, {Ahlers}, {Ahrens}, {Al Samarai},
  {Altmann}, {Andeen}, \& et~al.}]{icecube:18}
{IceCube Collaboration} {et~al.}, 2018, Science, 361, eaat1378

\bibitem[{{Koljonen} {et~al}\mbox{.}(2018){Koljonen}, {Maccarone},
  {McCollough}, {Gurwell}, {Trushkin}, {Pooley}, {Piano}, \&
  {Tavani}}]{koljonen18}
{Koljonen} K.~I.~I., {Maccarone} T., {McCollough} M.~L., {Gurwell} M.,
  {Trushkin} S.~A., {Pooley} G.~G., {Piano} G., {Tavani} M., 2018, \aap, 612,
  A27

\bibitem[{{Liska} {et~al}\mbox{.}(2018){Liska}, {Hesp}, {Tchekhovskoy},
  {Ingram}, {van der Klis}, \& {Markoff}}]{liska:18}
{Liska} M., {Hesp} C., {Tchekhovskoy} A., {Ingram} A., {van der Klis} M.,
  {Markoff} S., 2018, \mnras, 474, L81

\bibitem[{{Lister}(2016)}]{mojave16}
{Lister} M., 2016, Galaxies, 4, 29

\bibitem[{{McKinney}(2006)}]{mckinney:06}
{McKinney} J.~C., 2006, \mnras, 368, 1561

\bibitem[{{Merloni}, {Heinz} \& {di Matteo}(2003){Merloni}, {Heinz}, \& {di
  Matteo}}]{merloni:03}
{Merloni} A., {Heinz} S., {di Matteo} T., 2003, \mnras, 345, 1057

\bibitem[{{Motta}, {Casella} \& {Fender}(2018){Motta}, {Casella}, \&
  {Fender}}]{motta18}
{Motta} S.~E., {Casella} P., {Fender} R.~P., 2018, \mnras, 478, 5159

\bibitem[{{Narayan} \& {McClintock}(2012)}]{nmc12}
{Narayan} R., {McClintock} J.~E., 2012, \mnras, 419, L69

\bibitem[{{Russell}, {Gallo} \& {Fender}(2013){Russell}, {Gallo}, \&
  {Fender}}]{russellspin}
{Russell} D.~M., {Gallo} E., {Fender} R.~P., 2013, \mnras, 431, 405

\bibitem[{{Russell} {et~al}\mbox{.}(2013){Russell}, {Russell}, {Miller-Jones},
  {O'Brien}, {Soria}, {Sivakoff}, {Slaven-Blair}, {Lewis}, {Markoff}, {Homan},
  {Altamirano}, {Curran}, {Rupen}, {Belloni}, {Cadolle Bel}, {Casella},
  {Corbel}, {Dhawan}, {Fender}, {Gallo}, {Gandhi}, {Heinz}, {K{\"o}rding},
  {Krimm}, {Maitra}, {Migliari}, {Remillard}, {Sarazin}, {Shahbaz}, \&
  {Tudose}}]{russell13_maxi1836}
{Russell} D.~M. {et~al.}, 2013, \apjl, 768, L35

\bibitem[{{Tetarenko} {et~al}\mbox{.}(2019){Tetarenko}, {Casella},
  {Miller-Jones}, {Sivakoff}, {Tetarenko}, {Maccarone}, {Gandhi}, \&
  {Eikenberry}}]{tetarenko19}
{Tetarenko} A.~J., {Casella} P., {Miller-Jones} J.~C.~A., {Sivakoff} G.~R.,
  {Tetarenko} B.~E., {Maccarone} T.~J., {Gandhi} P., {Eikenberry} S., 2019,
  \mnras, 484, 2987

\bibitem[{{Tetarenko} {et~al}\mbox{.}(2018){Tetarenko}, {Freeman},
  {Rosolowsky}, {Miller-Jones}, \& {Sivakoff}}]{tetarenkoism}
{Tetarenko} A.~J., {Freeman} P., {Rosolowsky} E.~W., {Miller-Jones} J.~C.~A.,
  {Sivakoff} G.~R., 2018, \mnras, 475, 448

\bibitem[{{Tudor} {et~al}\mbox{.}(2017){Tudor}, {Miller-Jones}, {Patruno},
  {D'Angelo}, {Jonker}, {Russell}, {Russell}, {Bernardini}, {Lewis}, {Deller},
  {Hessels}, {Migliari}, {Plotkin}, {Soria}, \& {Wijnands}}]{tudor17}
{Tudor} V. {et~al.}, 2017, \mnras, 470, 324

\bibitem[{{van den Eijnden} {et~al}\mbox{.}(2018){van den Eijnden}, {Degenaar},
  {Russell}, {Wijnands}, {Miller-Jones}, {Sivakoff}, \& {Hern{\'a}ndez
  Santisteban}}]{eijnden18}
{van den Eijnden} J., {Degenaar} N., {Russell} T.~D., {Wijnands} R.,
  {Miller-Jones} J.~C.~A., {Sivakoff} G.~R., {Hern{\'a}ndez Santisteban} J.~V.,
  2018, \nat, 562, 233

\end{thebibliography}
}
%\end{thebibliography}

\end{document}